\documentclass{article} \usepackage{slashed,feynmf,epsfig,amsmath,amssymb,enumitem} \usepackage[papersize={8.5in,11in}]{geometry}
  \renewcommand{\vec}[1]{\boldsymbol{\mathrm{#1}}} \usepackage{bm}
\usepackage{graphicx}
\begin{document}
\title{Misaligned Spin Merging Black Holes in Modified Gravity (MOG)}
\author{J. W. Moffat\\
Perimeter Institute for Theoretical Physics, Waterloo, Ontario N2L 2Y5, Canada\\
and\\
Department of Physics and Astronomy, University of Waterloo, Waterloo,\\
Ontario N2L 3G1, Canada}
\maketitle


\begin{abstract}
A promising signature of coalescing black holes is their spin angular distribution. We consider the aLIGO collaboration gravitational wave measurements of the binary black hole spins and the predicted modified gravity (MOG) preference for misaligned spins of the coalescing black holes. In MOG, during the merger of two black holes, the enhanced strength of gravitation reduces the effective spin parameter $\chi_{\rm eff}\sim 0$ in agreement with the measured spin misalignment of the merging black holes observed in the gravitational wave detections by the aLIGO collaboration. 
\end{abstract}

\maketitle


\section{Introduction}

The direct detection of gravitational waves by the advanced LIGO collaboration~\cite{LIGO1,LIGO2,LIGO3,LIGO4} from merging black holes will allow for a determination of black hole formation environments. The angular distribution of black hole spins for systems formed through dynamical interactions in dense stellar environments is expected to have isotropic spins~\cite{PortegiesZwart,Rodriguez,Stone,Rodriguez2}, while black hole binaries formed from pairs of progenitor stars will preferentially have spins aligned with the orbital angular momentum~\cite{Mandel,Marchant}. A recent paper~\cite{Farr} investigated the possible angular distributions of the four binary black hole gravitational wave detections GW150914, LVT151012, GW151226 and GW170104 and their measured effective spins. The four detected events show a preference for the effective spin parameter $\chi_{\rm eff}$ to be clustered at zero. The observational data for low mass X-ray binaries spins show that the dimensionless spin parameters are in excess of $0.5$, and the spin alignments of the black holes can be affected by their formation during supernovae explosions~\cite{Miller}.

The effective dimensionless MOG spin parameter is given by~\cite{Moffat1}:
\begin{equation}
\label{MOGchi}
\chi_{\rm eff\,MOG}=\frac{c}{GM}\biggl(\frac{{\vec S}_1}{m_1}+\frac{{\vec S}_2}{m_2}\biggr)\cdot{\hat L}=\frac{m_1{\vec a}_1+m_2{\vec a}_2}{m_1+m_2}
\cdot{\hat{\vec L}},
\end{equation}
where in MOG~\cite{Moffat2} $G=G_N(1+\alpha)$ where $G_N$ is Newton's gravitational constant and $\alpha$ is a dimensionless scalar field which for the merging black holes is treated as a constant~\cite{Moffat1}. The $m_{1,2}$ are the gravitational masses of the binary components, $M=m_1+m_2$ is the total mass, ${\vec S}_{1,2}$ are the spin angular momentum vectors of the binary black holes and ${\hat L}$ is the unit orbital angular momentum vector normal to the angular momentum plane. The range of the effective spin parameter is: $-1 < \chi_{\rm eff} < 1$ corresponding to $-1$ for maximally misaligned spin and $+1$ for maximally aligned spins. The dimensionless spin magnitude parameter is
\begin{equation}
\label{spinmagnitudes}
a_{1,2}=\frac{c}{Gm_{1,2}^2}|{\vec S}_{1,2}|
\end{equation} 
for each black hole and is bounded by $0 < a_{1,2} < 1$.

\section{Kerr-MOG Black Hole}

An exact generalized Kerr solution of the Scalar-Vector-Tensor-Gravity (STVG) gravi-electric field equations has been derived~\cite{Moffat3,Moffat4,Moffat5,Armengol1,Armengol2}. The MOG field equations for $G=G_N(1+\alpha)\sim {\rm constant}$ and $Q_g=\sqrt{\alpha G_N}M\sim {\rm constant}$, where $G_N$ is the Newtonian gravitational constant and $\alpha$ is a dimensionless parameter  are given by
\begin{equation}
\label{phiFieldEq}
R_{\mu\nu}=-8\pi GT^\phi_{\mu\nu},
\end{equation}
\begin{equation}
\label{Bequation}
\frac{1}{\sqrt{-g}}\partial_\nu(\sqrt{-g}B^{\mu\nu})=0,
\end{equation}
\begin{equation}
\label{Bcurleq}
\partial_\sigma B_{\mu\nu}+\partial_\mu B_{\nu\sigma}+\partial_\nu B_{\sigma\mu}=0.
\end{equation}
We have ignored the small $\phi_\mu$ particle mass $m_\phi\sim 10^{-28}$ eV, $R_{\mu\nu}$ is the Ricci curvature tensor, $B_{\mu\nu}=\partial_\mu\phi_\nu-\partial_\nu\phi_\mu$ and $g={\rm Det}g_{\mu\nu}$.  The energy-momentum tensor ${{T^\phi}_\mu}^\nu$ is 
\begin{equation}
\label{Tphi}
{{T^\phi}_\mu}^\nu=-\frac{1}{4\pi}({B_{\mu\alpha}}B^{\nu\alpha}-\frac{1}{4}{\delta_\mu}^\nu B^{\alpha\beta}B_{\alpha\beta}).
\end{equation}

The Kerr-MOG black hole metric is given by
\begin{equation}
\label{KerrMOG}
ds^2=\frac{\Delta}{\rho^2}(dt-a\sin^2\theta d\phi)^2-\frac{\sin^2\theta}{\rho^2}[(r^2+a^2)d\phi-adt]^2-\frac{\rho^2}{\Delta}dr^2-\rho^2d\theta^2,
\end{equation}
where
\begin{equation}
\Delta=r^2-2GMr+a^2+\alpha(1+\alpha) G_N^2M^2,\quad \rho^2=r^2+a^2\cos^2\theta.
\end{equation}
Horizons are determined by the roots of $\Delta=0$:
\begin{equation}
r_\pm=G_N(1+\alpha)M\biggl[1\pm\sqrt{1-\frac{a^2}{G_N^2(1+\alpha)^2M^2}-\frac{\alpha}{1+\alpha}}\biggr].
\end{equation}
An ergosphere horizon is determined by $g_{00}=0$:
\begin{equation}
r_E=G_N(1+\alpha)M\biggl[1+\sqrt{1-\frac{a^2\cos^2\theta}{G_N^2(1+\alpha)^2M^2}-\frac{\alpha}{1+\alpha}}\biggr].
\end{equation}
The solution is fully determined by the Arnowitt-Deser-Misner (ADM) mass $M$ and spin parameter $a$ ($a=cS/GM^2)$. When $a=0$ the solution reduces to the Schwarzschild-MOG black hole metric:
\begin{equation}
\label{MOGmetric}
ds^2=\biggl(1-\frac{2G_N(1+\alpha)M}{r}+\frac{\alpha(1+\alpha) G_N^2M^2}{r^2}\biggr)dt^2-\biggl(1-\frac{2G_N(1+\alpha)M}{r}+\frac{\alpha(1+\alpha) G_N^2M^2}{r^2}\biggr)^{-1}dr^2-r^2d\Omega^2.
\end{equation}
The MOG solutions reduce to the general relativity (GR) Kerr and Schwarzschild black hole solutions when $\alpha=0$.

\section{MOG Effective Spin Parameter}

The inspiralling binary black holes initially agree with the GR phase for sufficiently separated black holes~\cite{Moffat1}, and only at the final merger of the black holes when the gravitational fields are strong and the black holes are moving at relativistic speeds will there be a deviation from GR. To derive the mass and spin parameters $M$ and $a$ for the merging black holes, it is necessary to solve the MOG field equations~\cite{Moffat1,Moffat2} in the manner of the numerical relativity solutions derived from the GR field equations~\cite{NumericalSolutions}. The numerical solutions of the MOG field equations will produce templates with different binary MOG black hole masses and spin parameters, which have to be matched to the measured aLIGO waveforms. However, we can obtain constraints on the spin alignments of the black holes, the magnitudes of spin parameter $a$ and the effective spin parameter $\chi_{\rm eff}$.

We treat the spin magnitudes $|{\vec S}_i|\,( i=1,2)$ of the binary black holes as measured quantities, while the dimensionless spin parameters $a_i$ are treated as inferred quantities 
given the spin values and the black hole masses. The spins of the binary black holes can be determined by post-Newtonian calculations of the inspiralling black holes, while the spins at the final merger phase can be determined by the frequency and time dependent waveforms obtained from the measured strain. A detection of $\sim 100$ gravitational wave events can be expected to allow for a determination of sufficiently accurate spin magnitudes. 

We obtain from Eq.(\ref{MOGchi}) the GR effective spin parameter for $\alpha=0$:
\begin{equation}
\chi_{\rm eff\,GR}=\frac{c}{G_NM}\biggl(\frac{{\vec S}_1}{m_1}+\frac{{\vec S}_2}{m_2}\biggr)\cdot{\hat L}.
\end{equation}
The ratio 
\begin{equation}
R_\chi\equiv\frac{\chi_{\rm MOG}}{\chi_{\rm GR}}=\frac{1}{1+\alpha}.
\end{equation}
For increasing $\alpha$, we have $R_\chi\rightarrow 0$ and MOG predicts a misalignment of the coalescing black holes spins that agrees with the measured spin misalignment orientation of the merging black holes for the four advanced LIGO gravitational wave detections. 

We see from Eq.(\ref{spinmagnitudes}) that the magnitudes of the black hole spin parameter $a$ will be reduced as $\alpha$ increases:
\begin{equation}
R_a\equiv\frac{a_{\rm MOG}}{a_{\rm GR}}=\frac{1}{1+\alpha}\rightarrow 0,
\end{equation}
which is consistent with the two right-hand contributions in Eq. (\ref{MOGchi}). As the two black holes merge the enhanced MOG gravitational strength $G=G_N(1+\alpha)$ for $\alpha > 0$  predicts $\chi_{\rm eff}\sim 0$ corresponding to a misalignment of the spins, and for sufficiently large $\alpha$ the prediction can agree with the average measured value $\chi_{\rm eff}\sim 0$. 

The MOG strong gravity prediction $\chi_{\rm  eff}\sim 0$ for $\alpha > 0$, signaling misaligned spins of the merging MOG black holes, introduces a constraint on the spin parameters that can play an important role in the final evolution phase of the inspiralling black holes. A preferred alignment of X-ray binary black hole spins in MOG is expected due to the large separation of the black holes with $\alpha\sim 0$, compared to the separation of binary black holes during the final merging phase when $\alpha > 0$. However, a natal kick experienced during the supernovae explosion formation of the black holes can misalign the spins~\cite{Miller}.

\section*{Acknowledgments}

I thank Luis Lehner, Martin Green and Viktor Toth for helpful discussions. This research was supported in part by Perimeter Institute for Theoretical Physics. Research at Perimeter Institute is supported by the Government of Canada through the Department of Innovation, Science and Economic Development Canada and by the Province of Ontario through the Ministry of Research, Innovation and Science.

\end{document}